\documentclass[aps,prb,showpacs,groupedaddress,twocolumn]{revtex4-1}
\usepackage{graphicx}
\usepackage{graphics}
\usepackage{amsmath,amssymb,amsfonts}
\begin{document}
\allowdisplaybreaks
%========================TITLE==========================
\title{The scaling feature of the magnetic field induced Kondo-peak splittings}

\author {Hui Zhang$^1$, X.C. Xie$^{1,2}$, and Qing-feng Sun$^{1,}$}
\altaffiliation{Electronic address: sunqf@aphy.iphy.ac.cn}
\address{
$^1$Institute of Physics, Chinese Academy of Sciences, Beijing
100190, China\\
$^2$Department of Physics, Oklahoma State University, Stillwater,
Oklahoma 74078 }
%===================ABSTRACT======================
\begin{abstract}

By using the full density matrix approach to spectral functions
within the numerical renormalization group method, we present a
detailed study of the magnetic field induced splittings in the
spin-resolved and the total spectral densities of a Kondo correlated
quantum dot described by the single level Anderson impurity model.
The universal scaling of the splittings with magnetic field is
examined by varying the Kondo scale either by a change of local
level position at a fixed tunnel coupling or by a change of the
tunnel coupling at a fixed level position. We find that the
Kondo-peak splitting $\Delta/T_K$ in the spin-resolved spectral
function always scales perfectly for magnetic fields $B<8T_K$ in
either of the two $T_K$-adjusted paths. Scaling is destroyed for
fields $B>10T_K$. On the other hand, the Kondo peak splitting
$\delta/T_K$ in the total spectral function does slightly deviate
from the conventional scaling theory in whole magnetic field window
along the coupling-varying path. Furthermore, we show the scaling
analysis suitable for all field windows within the Kondo regime and
two specific fitting scaling curves are given from which certain
detailed features at low field are derived. In addition, the scaling
dimensionless quantity $\Delta/2B$ and $\delta/2B$ are also studied
and they can reach and exceed 1 in the large magnetic field region,
in agreement with a recent experiment [T.M. Liu, et al., Phys. Rev.
Lett. 103, 026803 (2009)].
\end{abstract}

 \maketitle

\section{INTRODUCTION}

As the prototypical many-body phenomenon, the Kondo
effect\cite{ref1} has attracted much attention  for decades, and it
still occupies a central role for the understanding of many frontier
problems of condense matter physics. Dealing with the interaction
between a localized spin and delocalized conduction electrons, the
Kondo effect has a characteristic energy scale\cite{ref2,ref2.5}
$T_K$ which is defined as the Kondo temperature at which Kondo
conductance has decreased from its extrapolated zero-temperature
height to half of this value. By using $T_K$ as dimensionless unit,
many physical phenomena can exhibit universal scaling
relations\cite{ref3,ref4,ref5} which offer a shortcut to grasp the
intrinsic Kondo picture. One type of scaling analysis occurs when an
external magnetic field is applied to magnetic impurity systems (or
quantum dot (QD) \cite{ref6}): the splitting of zero-bias Kondo
conductance peak depends universally only on $B/T_K$ under a
magnetic field \cite{ref4,ref6,ref7,ref8,ref9,ref10,ref11}. That is
to say, in a wide magnetic field window, after rescaling treatment
all splittings under different parameters $\delta/T_K$ will follow
the same path with $B/T_K$ as the scaling variable. Such magnetic
field window can extend in theory from zero field up to fields $B\gg
T_K$,\cite{ref8,addref2,ref14} whereas in experimental
measurements\cite{ref6,ref12,ref26} magnetic fields rarely exceed
$5T_K$. For example, a detailed investigation in Kondo regime, in
particular for larger fields at around $100T_K$, has been made in
the work by Rosch et al.\cite{addref2} These scaling properties are
believed within the Kondo regime to be universal and cannot be
affected by system parameters, although such magnetic-field-induced
nonequilibrium effect has not reached to the level of the
equilibrium Kondo effect. Very recently, Liu et al\cite{ref12} found
large deviations from the conventional universal scaling analysis in
the measurement of Kondo differential conductance peak splitting
$\delta$ of a QD device. When adjusting $T_K$ in two different paths
(i.e. via energy level $\varepsilon_d$ or the coupling strength
$\Gamma$),  they found that the splitting $\delta$ also presented
two different trends with the increment of $T_K$ at a larger
magnetic field $B$, far beyond the traditional scaling
theory\cite{ref4,ref8,ref9,ref10} in Kondo effect. Such specific
experimental results indicate that the scaling characteristic may
encounter breakdown in certain situations, while all previous
theories\cite{ref4,ref8,ref9,ref10} indicate that scaling analysis
is independent on the path of adjusting $T_K$ even at a larger $B$.
Consequently it occurs to us that: does $T_K$ remain suitable as the
dimensionless scaling unit, especially in a larger magnetic field
situation? Therefore, a careful theoretical examination of the
scaling relation and its universal characteristic under an external
magnetic field is in great need.

In this paper, we present a detailed investigation of scaling in the magnetic
field induced splitting of the Kondo resonance in a Kondo QD, whose
universal characters are examined by adjusting $T_K$ in two
different paths (i.e.: via $\varepsilon_d$ and $\Gamma$). By
using the full density matrix numerical renormalization
group\cite{ref13} (FDM-NRG), we perform precise calculations of the
splitting of magnetic-field-induced Kondo peak in the total spectral
function (denoted as $\delta$) as well as in the spin-resolved
spectral function (denoted as $\Delta$). The FDM-NRG method has many
advantages compared to the conventional way used in previous
works\cite{ref9}, such as the
complete basis set, the
rigorous holding of the sum rules for the spectral function,
and less sensitivity
of the results to the number of kept states. Such new
features make our results more accurate than those from the previous
studies.

From the numerical results, we find that the splitting $\Delta$ in
the spin-resolved spectral function shows perfect universal scaling
characteristic no matter which path is taken to adjust $T_K$ as long
as the magnetic field $B$ is less than $8T_K$. For large fields $B$
(e.g. $B>10T_K$), deviations to scaling are found. For the
spin-averaged situation (corresponding to total spectral function),
small deviations of $\delta$ from universal scaling characteristic
always exist along the $\Gamma$-varying path as long as the magnetic
field is applied. Furthermore, we also fit the scaling curves
($f_{\delta}$ and $f_{\Delta}$) for both splittings $\delta/T_K$ (by
averaging out to eliminate small deviation) and $\Delta/T_K$. The
fitting curves $f_{\delta}$ and $f_{\Delta}$ have the following
characteristics: (1) The fitting curves $f_{\delta}$ has a threshold
value at about $0.5T_K$, while $f_{\Delta}$ is always non-zero under
a field $B/T_K$. (2) In the low field ($B\sim T_K$), $f_{\delta}$
exhibits the linear behavior with the slope coefficient $0.663$
which is consistent to $2/3$ from the Fermi liquid theory. (3) Two
curves $f_{\delta}$ and $f_{\Delta}$ get far away with one another and they do not
merge into a single curve in the large magnetic field region.  In
addition, the dimensionless scaling quantities $\Delta/2B$ and
$\delta/2B$ are also investigated and we find that they can reach
and exceed 1 when magnetic field is large enough, in agreement with
a recent experiment.\cite{ref12}

The rest of this paper is organized as follows. In Sec. \ref{NRG} we
introduce our model and the FDM-NRG method. In Sec. \ref{results} we
show numerical results of the magnetic-field-induced scaling
argument. Finally, Section \ref{conclusion} gives the conclusion.

\section{MODEL AND FDM-NRG METHOD}\label{NRG}

We consider the single-impurity Anderson model (SIAM)\cite{ref15}, and the Hamiltonian is given by:
\begin{eqnarray}\label{eq1}
H&=&H_{dot}+H_{leads}+H_{coupling}\nonumber\\
 H_{dot}&=&\sum\limits_{\sigma}\varepsilon_dd_{\sigma}^{\dagger}d_{\sigma}
 +Ud^{\dagger}_{\uparrow}d_{\uparrow}d^{\dagger}_{\downarrow}d_{\downarrow}\nonumber\\
H_{leads}&=&\sum\limits_{k\alpha\sigma}\varepsilon_{k\alpha}C_{k\alpha\sigma}^{\dagger}C_{k\alpha\sigma}\nonumber\\
H_{coupling}&=&\sum\limits_{k\alpha\sigma }t_{\alpha
}C_{k\alpha\sigma}^\dagger d_{\sigma}+h.c.
\end{eqnarray}
In the above Hamiltonian, the Fermion operator $C_{k\alpha\sigma}$
denotes the band states of leads with energy $\varepsilon_{k\alpha}$
and spin $\sigma$ ($\uparrow, \downarrow$), and $d_\sigma$ describes
the impurity states with energy $\varepsilon_d$; $U$ corresponds to
the Coulomb interaction between two electrons with different spin
embedded at the impurity site, and h.c. denotes ``hermitian
cojugate". Besides, $\alpha$ indicates different leads $L$ and $R$,
and $t_\alpha$ is the coupling between two subsystems. Accordingly,
the hybridization function between subsystems is given by:
\begin{align}
\Gamma_\alpha&=2\pi \sum\limits_k |t_\alpha|^2\delta(\omega-\epsilon_{k \alpha})\hspace{1em}(\alpha=L, R)\nonumber\\
\Gamma&=\Gamma_L+\Gamma_R
\end{align}
we can see that $\Gamma_{L/R}$, as well as $\Gamma$, will be constant in the broadband limits (adopted henceforth).

By using the canonical transformation:
\begin{eqnarray}\label{eq2}
a_{k\sigma}&=&\left(t_LC_{kL\sigma}+t_RC_{kR\sigma}\right)/\sqrt{t_L^2+t_R^2}\nonumber\\
b_{k\sigma}&=&\left(-t_RC_{kL\sigma}+t_LC_{kR\sigma}\right)/\sqrt{t_L^2+t_R^2}
\end{eqnarray}
we can see only the even combination of left and right electron
states $a_{k\sigma}$ couples to the local impurity state:
\begin{eqnarray}\label{q3}
H_{leads}&=&\sum_{k\sigma}\varepsilon_ka_{k\sigma}^\dagger a_{k\sigma}
+\sum_{k\sigma}\varepsilon_kb_{k\sigma}^\dagger b_{k\sigma}\nonumber\\ H_{coupling}&=&\sum_{k\sigma }Va_{k\sigma}^\dagger
d_{\sigma}+h.c.
\end{eqnarray}
where $V\equiv\sqrt{t_L^2+t_R^2}$.

There are many ways to solve SIAM, such as the slave-boson
mean-field theory, the Bethe ansatz approach, the numerical
renormalization group (NRG) method, {\it etc}, among which the Wilson's NRG
method\cite{ref16,ref17} has been proven to be an efficient and
powerful tool to deal with impurity system\cite{ref18}, especially
to obtain its Kondo features. If we are only interested in the
transport properties, there are standard procedures to make
use of NRG method:
\begin{itemize}
\item discretization of continuous Hamiltonian and its mapping to a semi-infinite chain;
\item iterative diagonalization of the chain and yield of flow of many-particle levels;
\item calculation of dynamic properties, such as the spectral functions.
\end{itemize}
The first two steps can be easily accomplished by following Ref. [\onlinecite{ref18}]
, and finally the Hamiltonian \eqref{eq1} becomes:
\begin{subequations}\label{eq3}
\begin{eqnarray}
H&=&\lim\limits_{n\rightarrow\infty}\Lambda^{-(n-1)/2}H_n\nonumber\\
 H_0&=&\Lambda^{-1/2}
\left(H_{dot}+\sum\limits_{\sigma}\sqrt{\frac{2\widetilde\Delta}{\pi}}\left(d_{\sigma}^\dagger
C_{0\sigma}+C_{0\sigma}^\dagger d_{\sigma}\right)\right)\hspace{8mm}\label{eq3a}\\
H_{n+1}&=&\sqrt{\Lambda}H_n+\Lambda^{n/2}\sum_{\sigma}
t_{n}\left(C_{n\sigma}^\dagger C_{n+1\sigma}+C_{n+1\sigma}^\dagger
C_{n\sigma}\right)\nonumber\\
&\label{eq3b}
\end{eqnarray}
\end{subequations}
where $\Lambda$ is the logarithmic discretization parameter,
$\widetilde\Delta\equiv\pi\sum\limits_kV^2\delta(\omega-\varepsilon_k)=
(\Gamma_L+\Gamma_R)/2 =\Gamma/2$ is also the hybridization function
between impurity subsystem and leads subsystem. $H_0$ is the
starting point of the above sequence of Hamiltonians Eq.
\eqref{eq3b} from which an iteration procedure can be established.
$t_n$, denoting the hopping term between two neighbor sites along
the chain, has an exponential decreasing feature with increasing $n$
and reduces to
$t_n\rightarrow\frac{1}{2}(1+\Lambda^{-1})\Lambda^{-n/2}$ in the
limit of large $n$.

Once Eq. \eqref{eq3} is obtained, there are many ways available to
derive dynamic properties by calculating impurity spectral function
in the Lehmann representation. Within the past few years, the
developments of NRG method, extending its application range to
various subjects including the bosonic\cite{ref19} and
time-dependent\cite{ref20} situations, are mainly in the subject of
dynamic properties. These successive improvements of NRG method, including
the reduced density matrix (DM)\cite{ref21}, the complete set of states combined with the reduced density matrix idea (CFS)\cite{ref22}, and most recently the full density matrix together the complete set of eliminated states (FDM)\cite{ref13}, have great
advantages compared to the conventional way\cite{ref2,ref23},
especially the last development FDM-NRG, which is far ahead of the
conventional method, exceeds in many aspects, such as dealing with
the impurity problem under an external magnetic field, describing
spectral features at finite temperature with high
accuracy, holding the sum rules of spectral function rigorously, using a complete basis set,
less sensitivity of the results to the number of kept states
, {\it etc}.

By using the FDM-NRG method, completeness relation reads as follows\cite{ref22}:
\begin{equation}\label{eq4}
1=\sum_{m=m_{min}}^N\sum_{l,e}|l,e\rangle_{m}^D\ _{m}^D\langle l,e|
\end{equation}
where $N$ denotes total length of the semi-infinite chain, $m_{min}$
is the first site at which states are discarded.
$|k,e\rangle_{m}^K=|k\rangle_m^K\otimes\{|\sigma_{m+1}\rangle
\}\otimes\ldots\otimes\{|\sigma_{N}\rangle\}$ and
$|l,e\rangle_{m}^D=|l\rangle_m^D\otimes\{|\sigma_{m+1}\rangle
\}\otimes\ldots\otimes\{|\sigma_{N}\rangle\}$, in which
$|k\rangle_m^K$ and $|l\rangle_m^D$ denote the kept and discarded
states of the $\sl{m}$th iteration shell respectively, and
$\{|\sigma_m\rangle\}$ represents the set of 4 states in the
$\sl{m}$th site along Wilson's chain (i.e.: $|0\rangle,
|\uparrow\rangle, |\downarrow\rangle, |\uparrow,
\downarrow\rangle$); thus $|k,e\rangle_{m}^K, |l,e\rangle_{m}^D$ can
be seen as kept and discarded states of the $\sl{m}$th shell
containing all information of whole system $H_N$ rather than only its first
$\sl{m}$ sites. Therefore, such treating technique naturally
involves the influence of ``environment"\cite{ref21}.

The next step is to make use of complete basis of set Eq.
\eqref{eq4} to solve the retarded Green's function which is given by:
\begin{equation} \label{eq5}
G_{AB}(t)=-i\Theta(t){\bf Tr} [\rho\{A(t),B(0)\}]
\end{equation}
where operators $A, B$ stand for $d_\sigma, d_\sigma^\dagger$
respectively, $\{A,B\}\equiv AB+BA$ is the standard anticommutation
relation, and $\rho$ is the full density matrix:
\begin{equation}\label{eq6}
\rho=\sum\limits_n\sum\limits_{le}|l,e\rangle_n^D\frac{e^{-\beta E_l^n}}{Z}\
_n^D\langle le|\equiv\sum\limits_nw_n\rho_n
\end{equation}
where $E_l^n$ is the $l$th (discarded) eigenstate of $n$th
iteration step, $\displaystyle w_n=\frac{4^{N-n}Z_n}{Z}$ is relative weight
holding the sum rule $\sum\limits_nw_n=1$, and
$\rho_n$ corresponds to the density
matrix of each single shell $n$: $\displaystyle(\rho_n)_{l,l^\prime}=\delta_{l,l^\prime}\frac{e^{-\beta
E_l^n}}{Z_n}$, in which
$Z_n=\sum\limits_l^De^{-\beta E_l^n}$ is its distribution function
containing only discarded states. Noticeably, such full density
matrix treatment, in contrast to the ``single-shell approximation" of
CFS-NRG method, is the spirit of FDM-NRG method.

Substituting the completeness relation Eq. \eqref{eq4} into the
retarded Green's function Eq. \eqref{eq5} and following the work by
Peter et al\cite{ref22}, we can easily derive final results in the
energy space as follows:
\begin{eqnarray}\label{eq7}
G_{AB}^r&=&G_{AB}^1+G_{AB}^2+G_{AB}^3\nonumber\\
G_{AB}^1&=&\sum\limits_n\left\{\
w_n\sum\limits_{l,l^\prime}A_{l,l^\prime}^{[DD]}B_{l^\prime,l}^{[DD]}
\cdot\frac{[\rho^{nn}]_{ll}+[\rho^{nn}]_{l^\prime l^\prime}}{\omega+E_l^n-E_{l^\prime}^n+i\eta}\right\}\nonumber\\
&+&\sum\limits_n\left\{\ w_n\sum\limits_{l,k}A_{l,k}^{[DK]}B_{k,l}^{[KD]}
\cdot \frac{[\rho^{nn}]_{ll}}{\omega+E_l^n-E_k^n+i\eta}\right\}\nonumber\\
&+&\sum\limits_n\left\{\ w_n\sum\limits_{l,k}B_{l,k}^{[DK]}A_{k,l}^{[KD]}\cdot
\frac{[\rho^{nn}]_{ll}}{\omega+E_k^n-E_l^n+i\eta}\right\}\nonumber\\
G_{AB}^2&=&\sum\limits_n\left\{
\sum\limits_{l,k,k^\prime}
\left[\sum\limits_{m>n}w_m[\rho^{nm}]_{k,k^\prime}\right]\cdot
\frac{A_{l,k}^{[DK]}\cdot B_{k^\prime,l}^{[KD]}}{\omega+E_l^n-E_k^n+i\eta}\right\}\nonumber\\
G_{AB}^3&=&\sum\limits_n\left\{
\sum\limits_{l,k,k^\prime}\left[\sum\limits_{m>n}w_m[\rho^{nm}]_{k^\prime,k}\right]
\cdot
\frac{B_{l,k^\prime}^{[DK]}\cdot A_{k,l}^{[KD]}}{\omega+E_k^n-E_l^n+i\eta}\right\}\hspace{4mm}
\end{eqnarray}
where $R_{rs}^{[XY]}\equiv\ _n^X\langle r|R|s\rangle_n^Y$, in which
$r, s\in\{k,l\}$ denote the kept ($k$) and discarded ($l$) states,
superscript $X, Y\in\{K, D\}$ indicates 'Kept' or 'Discarded', and
$R$ is the fermionic impurity operator $d_\sigma$ or
$d_\sigma^\dagger$. Such matrix elements can be obtained within
iteration process. Besides,
$[\rho^{mn}]_{k,k^\prime}\equiv\sum\limits_e\ _m^K\langle
k,e|\rho_n|k^\prime,e\rangle_m^K$ are elements of the $n$th
component of reduced density matrix of the shell $m$. Noticeably, Eq. \eqref{eq7} is an equivalent formulation of the FDM Green function which is identical to that in Ref. [\onlinecite{ref13}]. We
rewrite this result in a similar form as in the work by
Peter et al\cite{ref22} in order to reduce the time cost when
processing calculation; furthermore, Eq. \eqref{eq7} can be easily
reduced to the result in Ref. [\onlinecite{ref22}] by applying the
single-shell approximation $w_n=\delta_{n,N}$, thus it provides an
intuitional comparison between FDM-NRG method and CFS-NRG method in
the final result.

By using Eq. \eqref{eq7}, we can calculate the retarded Green's
function of impurity, and consequently all dynamic properties can be
obtained straightforwardly. For instance, the spectral
functions can be obtained at once:
\begin{equation}\label{eq8}
A_\sigma(\omega)=-\frac{1}{\pi}{\bf Im} G_{d_\sigma,d_\sigma^\dagger}^r
\end{equation}
such treatment yields a discrete spectral function rather than a
smooth one, and this problem can be overcome by using a broadening
function for each single $\delta$ peak. The kernel function in this
paper has a similar form as the one proposed by Weichselbaum and
Delft\cite{ref24}:
\begin{eqnarray}\label{eq9}
P_{LG}(\omega,\omega^\prime)&=&\frac{\theta(\omega\cdot\omega^\prime)}{\sqrt{\pi}\alpha|\omega|}\cdot\exp\left\{-\left(\frac{\log|\omega/\omega^\prime|}{\alpha}-\gamma\right)^2\right\}\nonumber\\
P_G(\omega,\omega^\prime)&=&\frac{1}{\sqrt{\pi}\beta}\cdot\exp\left\{-\frac{(\omega-\omega^\prime)^2}{\beta^2}\right\}\nonumber\\
H(\omega,\omega^\prime)&=&
\begin{cases}
1,\hspace{3.2cm}|\omega^\prime|>\omega_0\nonumber\\
\exp\left\{-(\frac{\log|\omega^\prime/\omega_0|}{\alpha})^2\right\},\ \ |\omega^\prime|\leq\omega_0
\end{cases}\nonumber\\
\delta(\omega-\omega^\prime)&\rightarrow&P_{LG}\cdot H+P_G\cdot(1-H)
\end{eqnarray}
where $\alpha, \beta, \gamma, \omega_0$ are relative parameters and
their values are fixed in practical calculation at a certain
temperature $T$: $\alpha=0.8, \beta=2T, \gamma=\alpha/4,
\omega_0=2T$. Noticeable, since $H(\omega, \omega^\prime)$ is
independent on parameter $\omega$, thus the above broadening
function make our spectral function hold sum rule identically on the
algorithms, also at finite  field $B$ and finite temperature $T$.

Since the broadening of $\delta$ function occurs only in the last
step of NRG method for spectral function calculation, thus the
influence of broadening on the splitting of Kondo-peak is actually
the approximation that we use a smooth distribution function to
replace the discrete $\delta$ function so that we can obtain smooth
spectral functions. There are many logarithmic features in Kondo
effect, thus the logarithmic Gaussian broadening function $P_{LG}$
is suitable for describing Kondo effect as well as certain features.
As the analysis in Ref. [\onlinecite{ref24}], the kernel broadening
function Eq. \eqref{eq9} can give the most accurate smooth spectral
functions than ever and it also has many advantages such as $P_{LG}$
is symmetric under $\omega \leftrightarrow \omega^\prime$ for the
choice $\gamma=\alpha/4$.

Noticeably, although we adopted FDM-NRG method\cite{ref13} here, it
is equivalent to CFS-NRG method\cite{ref22} in this paper,
because the investigation on magnetic-field-induced scaling analysis
in Kondo regime shown below is totally in the zero temperature case
where FDM-NRG method naturally becomes CFS-NRG method. Therefore,
the kernel broadening function Eq. \eqref{eq9} (i.e. : mixture of
Gaussian and logarithmic Gaussian) is not actually used in this
paper, but only the logarithmic Gaussian function $P_{LG}$ works.
Consequently the accuracy of our results is determined by $P_{LG}$,
and such accuracy has been enough to grasp Kondo picture and
describe precise Kondo features.

What's more, there is another technique to improve the accuracy: the
$z$ averaging\cite{ref18,ref25}, through which we can remove certain
oscillations in spectral functions. In this paper we also adopt
the $z$-averaging treatment in order to increase accuracy of our
results.

Finally, by using Eqs. \eqref{eq7}, \eqref{eq8} and \eqref{eq9} we
can calculate the impurity spectral function $A_\sigma(\omega)$
accurately and exactly, from which the Kondo physics can be derived
with great interest. Specifically these expressions are very helpful
for investigating the magnetic-field-induced scaling features shown
below.

Here we have to pinpoint one difference. In practical experiments
people usually measure how the splitting of Kondo
conductance peak, which is extracted in the conductance $G$ versus
the bias $V_{bias}$ curve, varies with the increment of magnetic
field, as done in the work by Liu et al\cite{ref12} for example.
By contrast, we investigate the magnetic-field-induced splitting in
spectral functions and keep the Fermi energy of left and right leads
aligned with each other, i.e.: $V_{bias}=0$. The two treatments are
equivalent to one another in studying the Kondo scaling features
under a magnetic field. Furthermore, the splitting of Kondo peak
in spectral functions is also experimentally measurable, e.g.,
by using an extra weak probe terminal\cite{addref1}.

\section{MAGNETIC-FIELD-INDUCED SCALING ARGUMENT}\label{results}

\begin{figure}
\includegraphics[width=\columnwidth,viewport=75 141 555 410,clip]{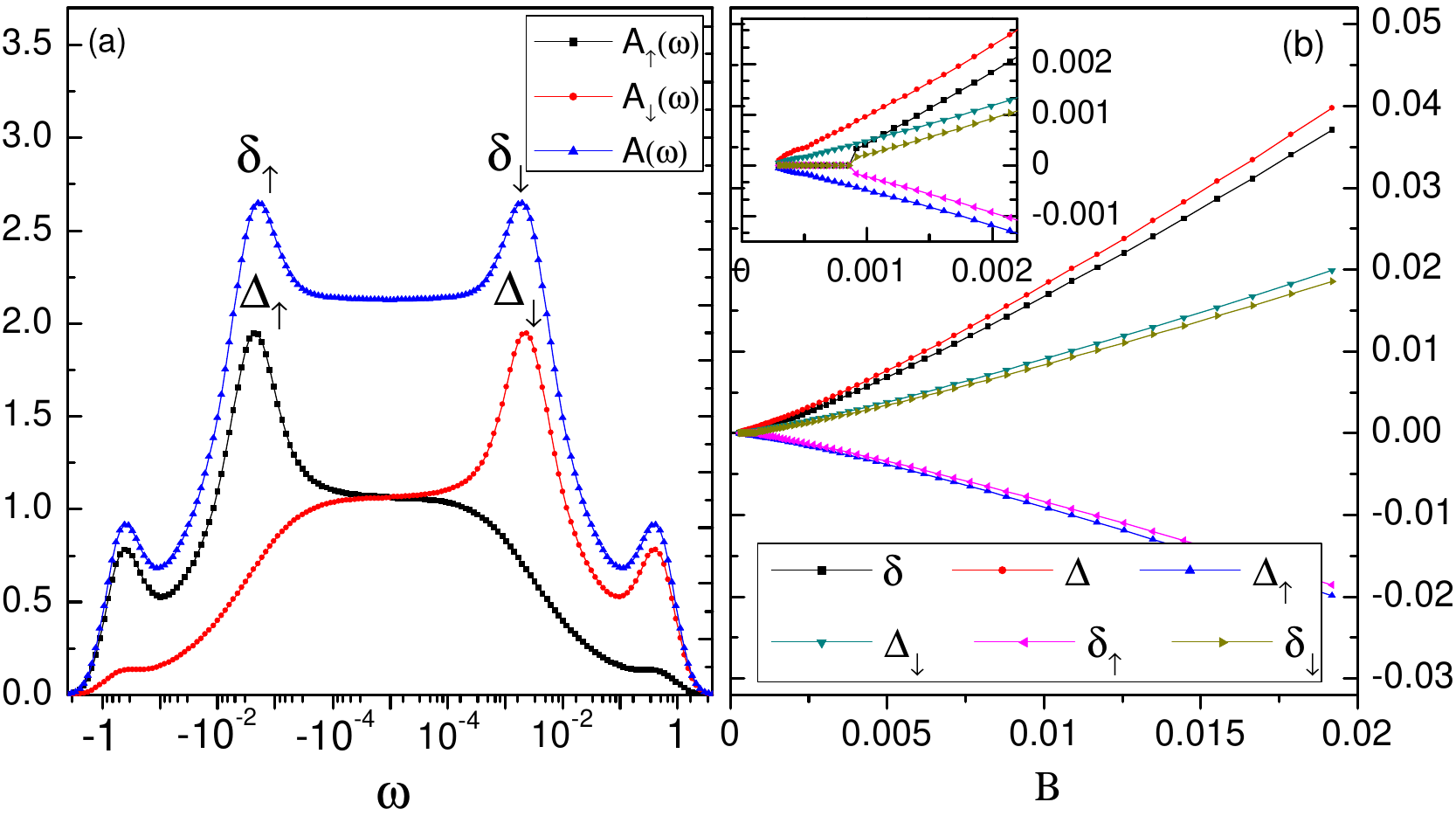}
\caption{(Color online) (a) The spin-resolved spectral functions
$A_{\uparrow}(\omega)$ and $A_{\downarrow}(\omega)$ and the total
spectral functions $A(\omega)$ versus the energy $\omega$ at the
magnetic field $B=3.273\times10^{-3}$. (b) shows the changes of the Kondo peak
positions $\delta_\uparrow$, $\delta_\downarrow$, $\Delta_\uparrow$,
and $\Delta_\downarrow$ and the splittings $\delta$ and $\Delta$
with the increment of magnetic field $B$, with the inset as its
magnification in a low magnetic field window. The parameters of SIAM
for (a) and (b) are the same: $U=1, \varepsilon_d=-0.5$, and
$\Gamma=0.16$ with its kondo temperature
$T_K=1.476\times10^{-3}$; NRG parameters: $\Lambda=2.5, Ns=150$, and
$T=0$.} \label{figure1}
\end{figure}

We calculate the spin-resolved spectral functions
$A_\uparrow(\omega), A_\downarrow(\omega)$ and total spectral
function $A(\omega)$ by using Eqs. \eqref{eq7}, \eqref{eq8} and
\eqref{eq9} of FDM-NRG method. In equilibrium situation without an
applied magnetic field, there are three peaks coming out in spectral
functions, among which two peaks correspond to the energy level at
$\omega=\varepsilon_d, \varepsilon_d+U$, while the third one
corresponds to Kondo resonance peak roughly at the Fermi level
$\omega=0$. Under an external magnetic field $B$, however, the Kondo
resonance peak will splits into two separate peaks as shown in Fig.
\ref{figure1}(a). These peak positions, denoted as $\delta_\uparrow,
\delta_\downarrow$ in the spin-averaged case and $\Delta_\uparrow,
\Delta_\downarrow$ in the spin-resolved case, can be extracted from
spectral functions at the maximum values which can be determined by
analyzing derivatives. Through scanning a certain region, which
contains two stagnation points related to the splittings, by using
binary search method, the splitting positions can be localized with
high precision. In our calculation, this precision reaches
$0.001T_K$. The splittings $\delta$ and $\Delta$ are then given by
$\delta=\delta_\downarrow-\delta_\uparrow$ and
$\Delta=\Delta_\downarrow-\Delta_\uparrow$. In the particle-hole
symmetry case (i.e.: $\varepsilon_d=-U/2$), these peaks are symmetry
corresponding to the Fermi energy $\omega=0$. Fig.\ref{figure1}(b)
shows these peak positions ($\delta_{\uparrow}$,
$\delta_{\downarrow}$, $\Delta_{\uparrow}$, and
$\Delta_{\downarrow}$) and the splittings ($\delta$ and $\Delta$) as
the function of magnetic field $B$. From Fig.\ref{figure1}(b), we
can see three important magnetic-field-induced features: First, the
splitting $\delta$ has an obvious threshold value which has been
predicted by previous theories\cite{ref8,ref9}, whereas the
splitting $\Delta$ in the spin-resolved case always exists even in a
very weak magnetic field $B$. Second, the $\delta$-$B$ curve can be
seen as linear in a low field window while the $\Delta$-$B$ curve
deviates from this feature. Third, with increasing field $B$, the
two curves do not get closer to each other, conversely
they become far away from one another. These features will be
investigated in detail below.

\begin{figure}
\includegraphics[width=\columnwidth,viewport=48 0 438 453,clip]{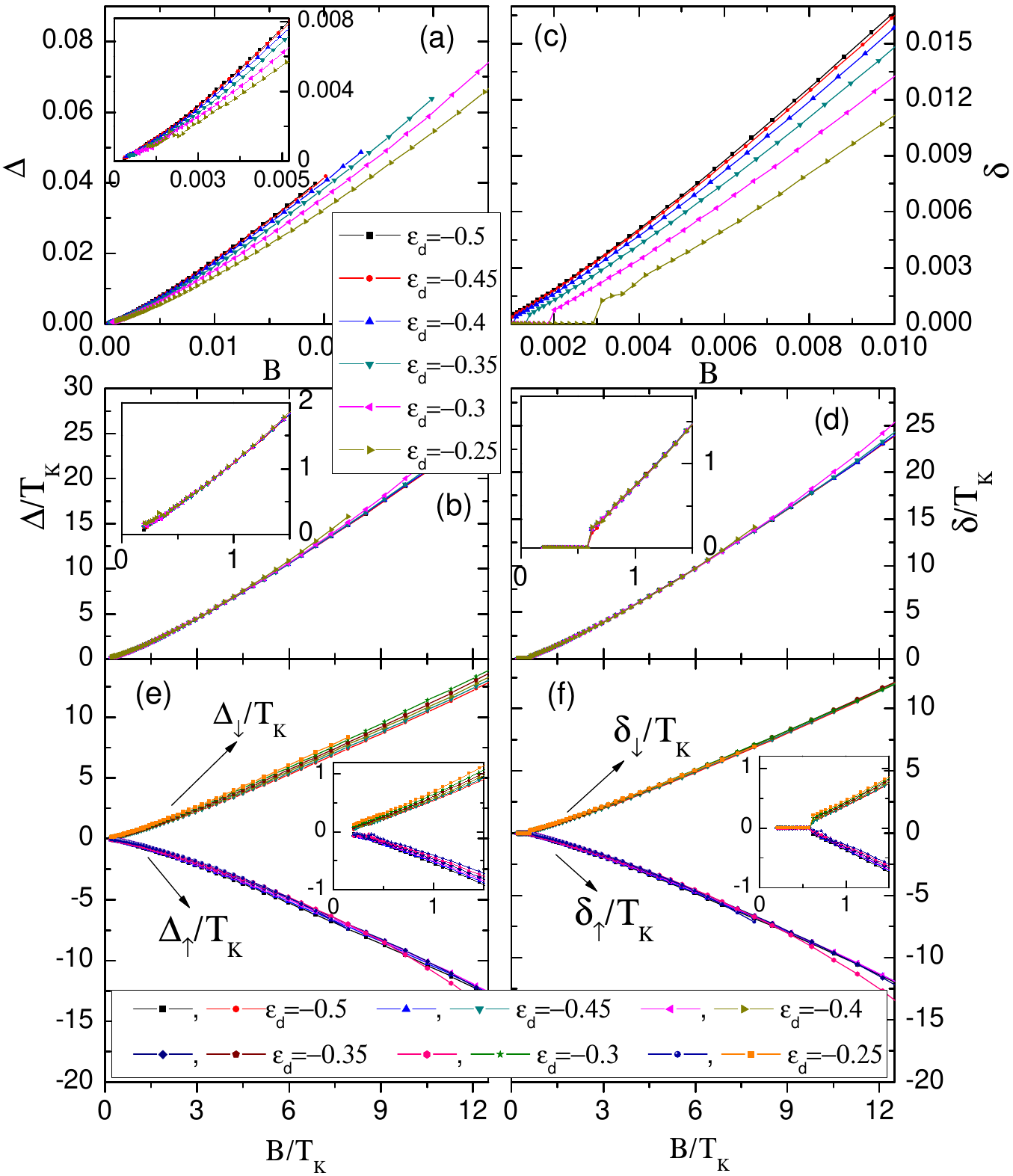}
\caption{(Color online) (a) and (c) represents the splittings
$\Delta$ and $\delta$ vs $B$ with different $\varepsilon_d$ in the
spin-resolved and total spectral function respectively. (b) and
(d) are the results of (a) and (c) after scaling treatment. The
splitting positions in the spin-resolved spectral function with
different $\varepsilon_d$ are presented in (e), while such positions
in the spin-averaged case are shown in (f). All insets are
magnifications in a low $B$ window to show certain details. All
parameters are the same as in Fig. \ref{figure1} except the energy
level $\varepsilon_d$ varying from $-0.5$ to $-0.25$, with its Kondo
temperature varying from $1.476\times10^{-3}$ to
$5.037\times10^{-3}$.} \label{figure2}
\end{figure}

We first find out how the scaling curves change with varying energy
level $\varepsilon_d$ which is one path of adjusting
Kondo scale\cite{haldane}:
\begin{equation}
T_K=\frac{1}{2}\sqrt{\Gamma U}\exp[\pi \varepsilon_d(\varepsilon_d+U)/\Gamma U]
\end{equation}
where $\Gamma$ is the whole width of impurity's energy level. In
Fig. \ref{figure2} (a) and (c), we can see the raw data extracted
directly from spectral functions changing gradually along
parameter-dependent paths, either in the spin-resolved case or in
the spin-average case. It means that if we change some parameters,
the evolution of the splittings $\Delta$ and $\delta$ vs $B$ will
go along another different path at once. However, after scaling
treatment by using Kondo scale $T_K$ as the dimensionless unit,
these changes of all splittings with increasing $B$ surprisingly
go along the same path, as seen in Fig. \ref{figure2} (b) and (d).
Such feature, which is called the scaling characteristic of Kondo
effect, is the main subject under detailed investigation in this
paper. From the results of Fig.2 (b) and (d), the scaling
characteristic will work well for magnetic field $B<8 T_K$.
Whereas with magnetic field getting larger some deviations appear,
because the system has been driven closer to the Kondo regime in a
larger $B$ region. When the magnetic field $B$ is larger than $10
T_K$, in which Kondo effect has been suppressed badly and the QD
system is out of Kondo regime, obvious deviations in both of
splittings $\delta$ and $\Delta$ are exhibited and the scaling
theory won't be obeyed.

\begin{figure}
\includegraphics[width=\columnwidth,viewport=66 28 517 409,clip]{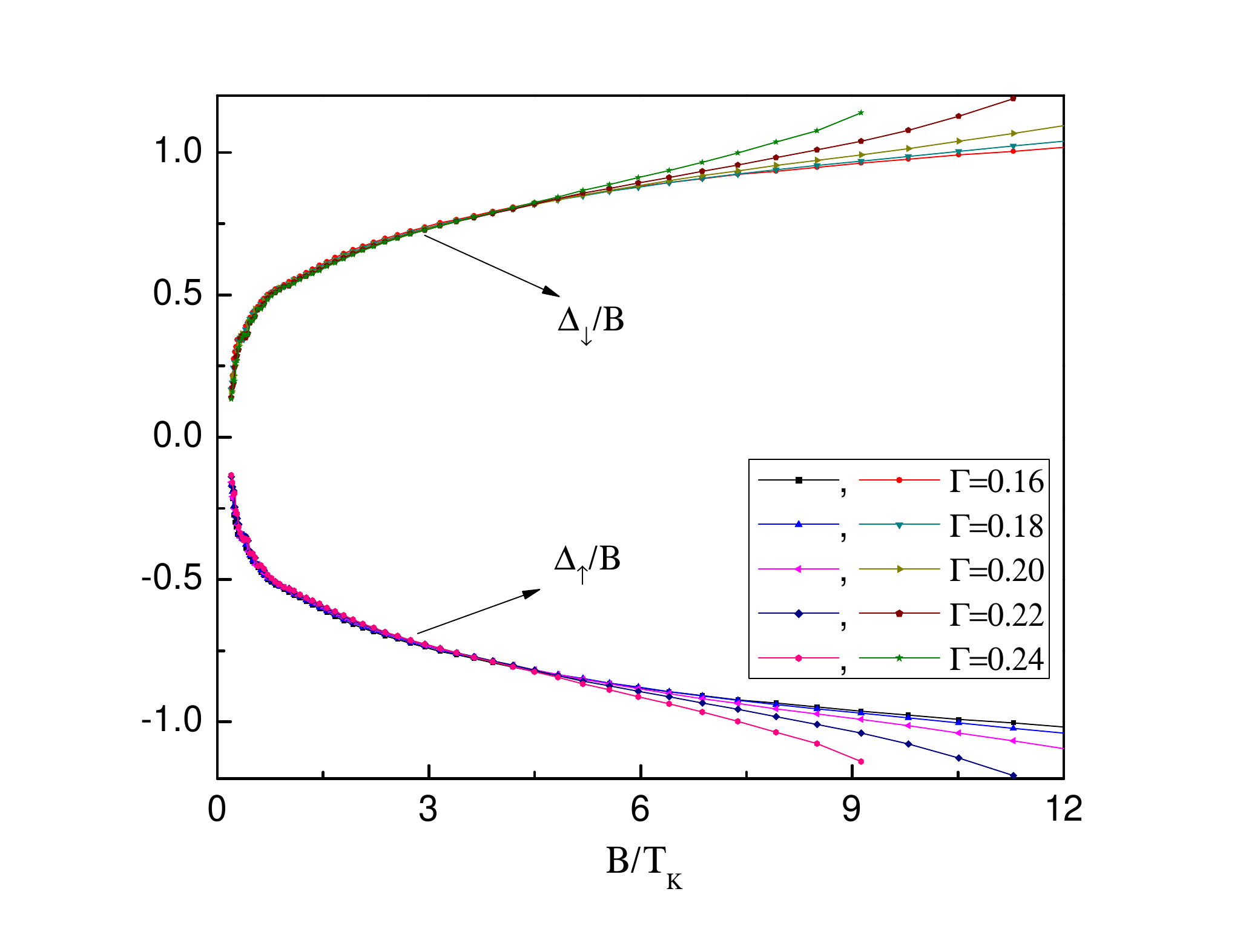}
\caption{(Color online) changes of $\Delta_\uparrow/B$ and $\Delta_\downarrow/B$ with the increment of $B/T_K$ by varying $\Gamma$. The parameters are the same as in
Fig. \ref{figure1} except the coupling strength $\Gamma$ varying
from $0.16$ to $0.24$, with its Kondo temperature varying from
$1.476\times 10^{-3}$ to $9.286\times 10^{-3}$.} \label{figadd}
\end{figure}

In order to clearly show that the system indeed is out of Kondo
regime when the magnetic field $B$ is larger than $10 T_K$, we plot
the curves $\Delta_{\uparrow}/B$ and $\Delta_{\downarrow}/B$
versus $B/T_K$ by varying $\Gamma$ (see Fig. \ref{figadd}). Here it can be
seen clearly that the deviation (i.e.: the non-universal features)
are pushed to much larger field $B$ with decreasing $\Gamma$, and in
addition no deviations from scaling are found in a low field $B$.
As a result, the deviations from scaling found at larger field
$B>10T_K$ are indeed due to leaving the Kondo regime. In the
following, we will mainly focus on the Kondo regime with the
magnetic field $B$ less than $10 T_K$.

Under further investigation of the scaling feature, two differences
between the spin-resolved case and spin-averaged case reveal. At
first the splitting $\delta$ has an obvious threshold field value at
around $0.5T_K$, by contrast we can't see such threshold
field in the $\Delta/T_K$ vs $B/T_K$ curve. This is because the
two separated peaks in $A_\uparrow(\omega)$ and
$A_\downarrow(\omega)$ are too close to each other that they overlap
and can not be separated in total spectral function. The second
difference is about the scaling feature. From (b) and (d) (or by
comparison between the insets) we can see the $\delta/T_K$ vs
$B/T_K$ curve presents an linear scaling characteristic, whereas the
$\Delta/T_K$ vs $B/T_K$ curve shows a nonlinear behavior in a low
field window.

Next we study the scaling characteristics for the positions of the
Kondo peaks $\Delta_{\uparrow}$, $\Delta_{\downarrow}$,
$\delta_{\uparrow}$, and $\delta_{\downarrow}$. Fig.2 (e) and (f)
show these peak positions ($\Delta_{\sigma}/T_K$ and
$\delta_{\sigma}/T_K$) vs the magnetic field $B /T_K$ at
different energy level $\varepsilon_d$. In approximation, all curves can
merge to one curve and the scaling characteristics can hold.
However, in detail, all positions of the Kondo peak slightly shift
towards the same direction by varying energy level, in the whole
magnetic field region (including the weak magnetic field $B<T_K$).
The deviations are within $0.5 T_K$ and the deviations in the
$\Delta_{\sigma}/T_K$-$B/T_K$ curves are larger than that in the
$\delta_{\sigma}/T_K$-$B/T_K$ curves. In addition, due to the
deviations of the $\Delta_{\sigma}$ and $\delta_{\sigma}$ are
towards the same direction, the splittings $\Delta$
($\Delta=\Delta_\downarrow-\Delta_\uparrow$) and $\delta$
($\delta=\delta_\downarrow-\delta_\uparrow$) can reduce certain
parameter-induced deviations and well keep the scaling feature.

\begin{figure}
\includegraphics[width=\columnwidth,viewport=80 0 530 453,clip]{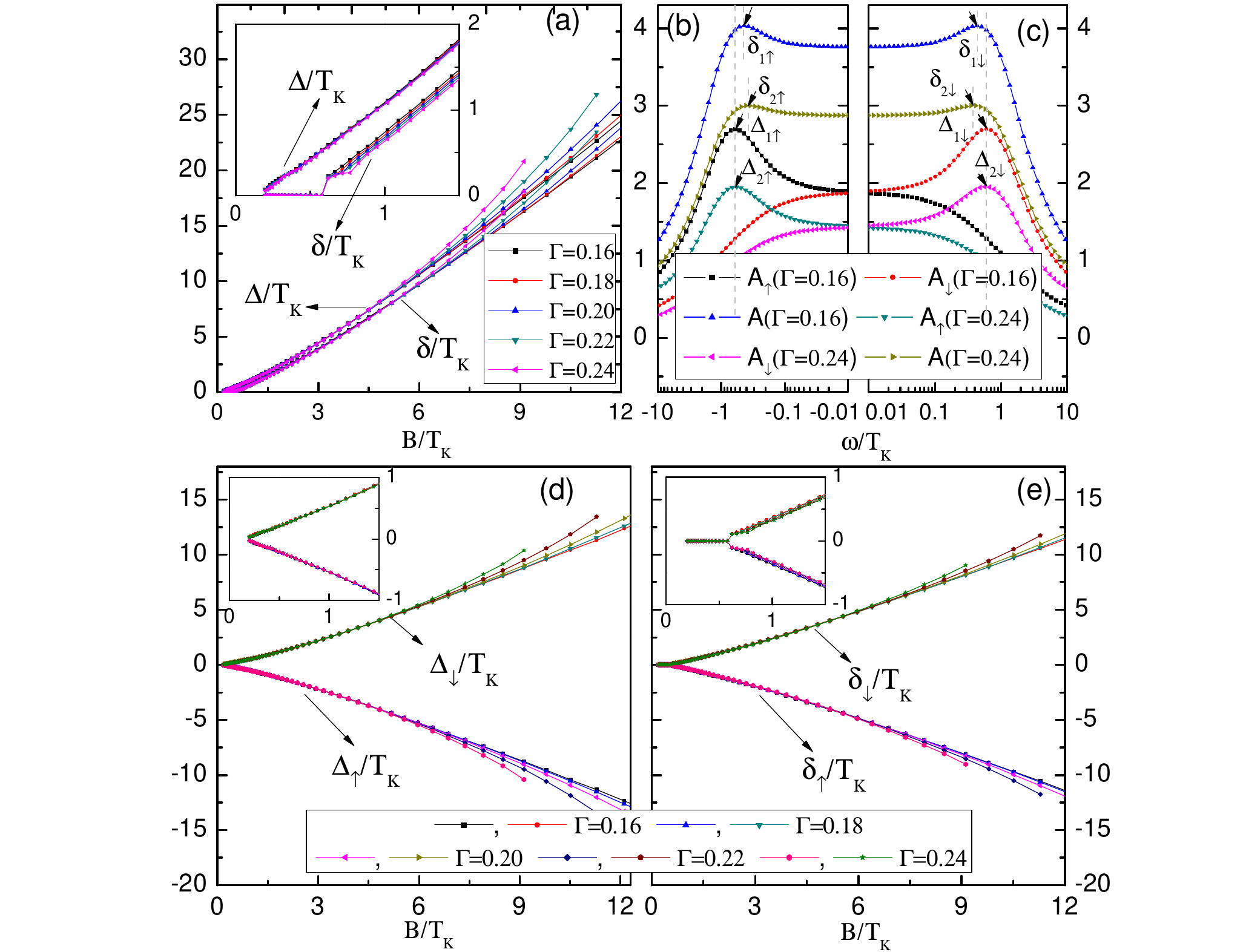}
\caption{(Color online) The variations of splittings $\Delta/T_K$
and $\delta/T_K$ vs $B/T_K$ for different coupling strength $\Gamma$
are exhibited in (a), and the corresponding Kondo-peak positions
$\Delta_{\sigma}/T_K$ and $\delta_{\sigma}/T_K$ in the
spin-resolved case and in the spin-averaged case are shown in (d)
and (e) respectively. In addition (b) and (c) present the total and
spin-resolved spectral functions for two specific cases:
$\Gamma=0.16$ and $\Gamma=0.24$. All parameters are the same as in
Fig. \ref{figadd}.} \label{figure3}
\end{figure}

Following we adjust Kondo scale $T_K$ in another path by varying the
coupling strength $\Gamma$, in order to check whether scaling
characteristic also works well in this situation. We find out some
unusual details reveal in the inset of Fig. \ref{figure3} (a)
although the behavior of splittings with increasing magnetic field
shown in Fig. \ref{figure3} (a) looks much similar like the one in
Fig. \ref{figure2} (b) and (d). Roughly speaking, we can consider
the scaling feature as an effective characteristic yet. For the
spin-resolved case, ignoring the larger field region where system
has been driven closer to or out of Kondo regime, the filed-induced
splitting $\Delta/T_K$ scales perfectly for magnetic fields
all the time. But for the spin-averaged case, we can't ignore such a
fact that evolution of splittings $\delta/T_K$ with the increment of
$B/T_K$ don't go along exactly the same path, as is seen in inset of
(a). Furthermore, we find that such slight deviation always exists
as long as the magnetic field is applied in system, and a larger
field window is not necessary to reveal this feature along the
$\Gamma$ varying path.

The reason for inducing such deviation can be seen in Fig.
\ref{figure3} (b) and (c) where the spectral functions under two
specific coupling strength $\Gamma=0.16$ and $0.24$ (denoted by
using subscripts $1$ and $2$ respectively) are shown. Since the
magnetic field is the only factor inducing Kondo peak splitting as
mentioned above, the splitting positions $\Delta_\uparrow/T_K$ and
$\Delta_\downarrow/T_K$ should be dependent only on field $B$ and
$T_K$. Along the $\varepsilon_d$ varying path, we have seen that
$\Delta/T_K$ scales perfectly, and here we see this effect also
works well along the $\Gamma$ varying path: as shown in
Fig.\ref{figure3} (b) and (c), the positions $\Delta_{\sigma}/T_K$
of the Kondo peak under two different coupling strength are indeed
nearly the same. On the other hand, the positions
$\delta_{\uparrow(\downarrow)}/T_K$ in the spin-averaged case are
not so luck like the one in spin-resolved case. Certain parameters
can really influence specific shape of spectral function and the
high of Kondo peak (although the Kondo-peak position keeps fixed),
and this makes the peak positions of total spectral functions
shift a little towards Fermi energy (i.e.: $\omega=0$) with
increasing coupling strength $\Gamma$ as shown in
Fig.\ref{figure3} (b) and (c). Here is in the place to pinpoint
that such effect occurs as long as $B$ is applied, that's to say
the deviation from conventional scaling analysis will be observed
under any value of external magnetic field, although this
deviation is usually small (within $0.2T_K$ in our parameters, see
the inset of Fig.\ref{figure3} (a)). This result is consonant with
the experimental work by Liu et al\cite{ref12} in some aspects
that the coupling strength can indeed induce deviation from
scaling theory, but moreover our work indicates that such
deviation occurs all the time along $\Gamma$ varying path, rather
than localized in a larger field window.

Another possible way, which explains why $\Gamma$ can induce
the small deviation from scaling theory, is to investigate Kondo
peak positions shown in Fig. \ref{figure3} (d) and (e). In present
situation, the system remains symmetry ($2\varepsilon_d+U=0$) and
thus these peak positions keep symmetry as well corresponding to the
Fermi energy $\omega=0$. The Kondo-peak positions
$\Delta_{\sigma}/T_K$ in the spin-resolved case keep nearly the
same with changing $\Gamma$ while the magnetic field $B<6T_K$, thus
$\Delta/T_K$-$B/T_K$ curve fits the scaling theory well in this
field window. In the spin-averaged case, however, the peak position
$\delta_{\sigma}/T_K$ slightly deviates with changing $\Gamma$. Now
the deviation of $\delta_{\uparrow}/T_K$ and
$\delta_{\downarrow}/T_K$ are in the oppositive direction because of
the symmetrical peak positions. So this small deviation are not
eliminated but enlarged in the splitting $\delta/T_K$ and the scaling theory can
not hold exactly along the $\Gamma$ varying path.

Since observable deviation of $\delta/T_K$ vs $B/T_K$ from scaling
theory is usually small as is seen in Fig. \ref{figure3} (a), then
such small deviation can be eliminated by averaging out so that
scaling analysis can be processed. Next we present a scaling
analysis suitable for all field windows within the Kondo regime, for
both the spin-resolved case and spin-averaged case.

\begin{figure}
\includegraphics[width=\columnwidth,viewport=80 9 554 446,clip]{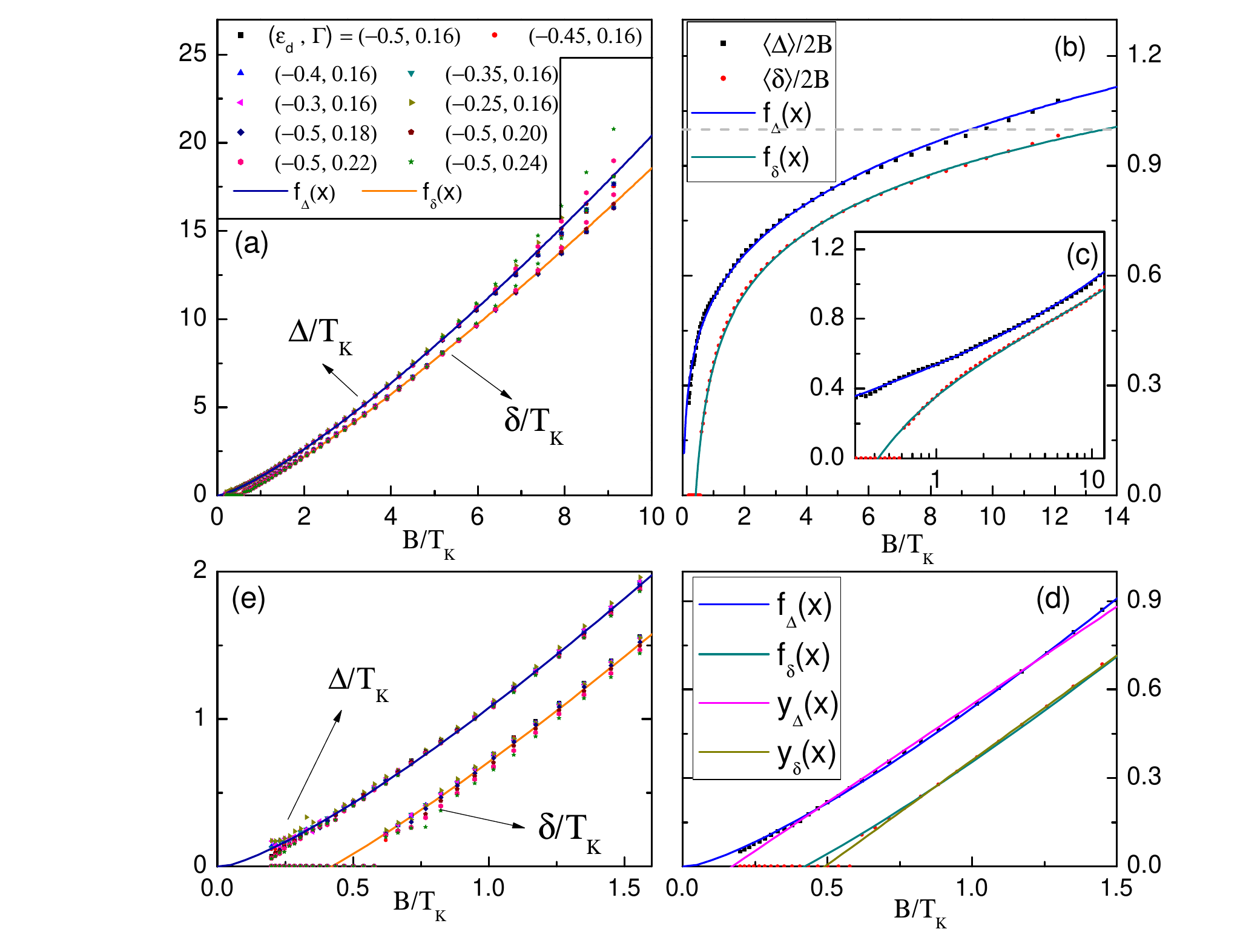}
\caption{(Color online) The comparison between the fitting curves
$f_\Delta$ and $f_\delta$ and numerical data under various
parameters are exhibited in subfigure (a), and (e) is its
magnification in a low field window. (b) presents the fitting curves
$f_{\Delta}/2B$ and $f_{\delta}/2B$ and the dimensionless average
data $\langle\Delta\rangle/2B$ and $\langle\delta\rangle/2B$, where
$\langle\Delta\rangle$ and $\langle\delta\rangle$ are the average
values corresponding to the raw numerical data in (a). The inset (c)
is the same with (b) but under the logarithmic coordinate. (d)
presents $f_{\Delta}$, $f_{\delta}$, $y_{\Delta}$ and $y_{\delta}$
while the magnetic field $B$ is in the vicinity of $T_K$. The other
parameters are the same as in Fig. \ref{figure1}} \label{figure4}
\end{figure}

In Fig. \ref{figure4} (a) we exhibit the comparison between fitting
curves and raw data extracted directly from the spectral functions.
Such fitting curves are given by
\begin{equation}
\begin{aligned}
f_\Delta(x)&=a x^{b}\exp(-c/x)\\
f_\delta(x)&=d x\exp(-f/x^{0.1})+h
\end{aligned}
\end{equation}
with the parameters: $a=1.1185, b=1.2627, c=0.03992; d=18.2913,
f=2.8572, h=-0.3421$. From Fig. \ref{figure4} (a) we can see the
fitting curves fit numerical data very well in whole magnetic
window, including a low field below the conventional threshold value
$0.5T_K$. Thus we can study some detailed features in a low field
with the help of fitting curves. For example, the splitting $\delta$
in spin-averaged case does have a threshold magnetic field at around
$0.5T_K$ as is seen in (e), and this result is in quantitative
agreement with previous theories\cite{ref9}. What's more, by using
fitting curve $f_\Delta(x)$ we find out that in a low magnetic
field the curve $\Delta/T_K$ vs $B/T_K$ likes the power
function and the first derivative of $\Delta/T_K$ with respect to
$B/T_K$ is always zero at the zero-field point. When magnetic
field $B$ is in the vicinity of $T_K$, the curves $f_\Delta(x)$ and
$f_\delta(x)$ exhibit the linear behaviors:
\begin{equation}
\begin{aligned}
y_\Delta(x)&=a_1x+b_1\\
y_\delta(x)&=a_2x+b_2
\end{aligned}
\end{equation}
with the parameters: $a_1=0.6632, b_1=-0.1136; a_2=0.7097,
b_2=-0.3492$. In Fig. \ref{figure4} (d), we show the curves
$f_\Delta(x)$, $f_\delta(x)$, $y_\Delta(x)$, and $y_{\delta}(x)$
together. It clearly shows that $y_{\delta/\Delta}(x)$ is well
consistent with $f_{\delta/\Delta}(x)$. In particular, the
coefficients $a_1$ of the slope of the curve $y_{\delta}(x)$ is
$0.6632$, which is nearly the same as $2/3$ derived through Fermi
liquid theory.\cite{ref11}

On the other hand, in the high magnetic field region, the two curves
$f_{\Delta}(x)$ and $f_{\delta}(x)$ do not close, whereas they get
far away with the increment of $B$. This result is quite surprise.
Generally speaking it is believed that splittings $\Delta/T_K$ and
$\delta/T_K$ in the spin-resolved and spin-averaged cases should
reach the limit of Zeeman splitting  under a large enough
magnetic field, thus the two scaling curves $f_{\Delta}(x)$ and
$f_{\delta}(x)$ should get closer to one another rather than far
away with increasing magnetic field. The key point for the getting
far away behavior is that the peak position $\Delta_{\uparrow}$ always falls
in the region where its second derivative of spin-down spectral
function $A_{\downarrow}$ with respect to $\omega/T_K$ is negative
(see Fig. \ref{figure3} (b) and (c)).

In Fig. \ref{figure4} (b) (c) and (d) we show other scaling features
investigated by previous works.\cite{ref8,ref9} $\delta/2B$ is
another scaling dimensionless quantity and has been predicted to
reach the limit value $1$ (i.e. the splitting of the Kondo peak is
equal to the Zeeman splitting) in the large magnetic field $B$
limits.\cite{ref8,ref9} Whereas in recent experiments, such the
limit value can be exceeded\cite{ref7,ref12}. In Fig. \ref{figure4}
(b), we show the evolution of $\delta/2B$ vs the magnetic field
$B/T_K$. The numerical results clearly exhibit that $\delta/2B$ can
reach and exceed 1, which is in agreement with the recent
experiments\cite{ref7,ref12}. In addition, in the large field
window, the curve of $\delta/2B$-$B/T_K$ shows the logarithmic
behavior.\cite{ref10} This behavior can clearly be seen in Fig.
\ref{figure4} (c), in which the curve of $\delta/2B$-$ln(B/T_K)$ is
shown.

\section{CONCLUSION}\label{conclusion}

In conclusion, by using the FDM-NRG method we study the scaling
characteristics of the Kondo-peak splittings in a quantum dot system
under a magnetic field. Similarly as in the recent
experiment,\cite{ref12} two different ways to adjust the Kondo
scale $T_K$, via the energy level $\varepsilon_d$ and the
coupling strength $\Gamma$, are considered. Both splittings $\Delta$
and $\delta$ of Kondo resonant peaks in the spin-resolved
spectral function and the total spectral function are investigated
in detail. We find that the splitting $\Delta/T_K$ in the
spin-resolved case always scales perfectly for magnetic fields $B<8T_K$ and regardless of
the $T_K$-adjusted paths. When the magnetic field $B$ is over
$10T_K$, obvious deviations are exhibited since QD system has been
out of the Kondo regime. On the other hand,
$\delta/T_K$ in the total spectral function, which can be related to the experiments by Liu et al\cite{ref12}, deviates slightly from
the conventional scaling theory along the $\Gamma$-varying path. Such result is consonant with the experimental work by Liu et al in some aspects, i.e.: the coupling strength can indeed induce deviation from scaling theory, but moreover our work indicates that such deviation occurs all the time as long as an external magnetic field is applied. Therefore $T_K$ is an
unsuitable parameter as the scaling dimensionless unit in this
situation. Since the deviation in the splitting $\delta/T_K$ is
usually small, we can still make the scaling analysis for
$\delta/T_K$. The fitting curves ($f_{\delta}$ and $f_{\Delta}$) for
both splittings $\delta/T_K$ and $\Delta/T_K$ are presented. The
fitting curves $f_{\delta}$ has a threshold value at around $0.5T_K$,
but $f_{\Delta}$ is always non-zero while under the field $B/T_K$.
In the low field ($B\sim T_K$), $f_{\delta}$ exhibits the linear
behavior with its slope coefficient $0.663$, consistent with the
value of $2/3$ from the Fermi liquid theory. On the large field
side, two curves $f_{\delta}$ and $f_{\Delta}$ get apart, and
both $\Delta/2B$ and $\delta/2B$ can reach and exceed 1, which is also in
agreement with a recent experiments.

\section{ACKNOWLEDGMENTS}
We gratefully acknowledge Ning. Hua. Tong for helpful discussion on
the NRG method. This work was financially supported by NSF-China
under Grants Nos. 10734110, 10821403, and 10974236, China-973
program and US-DOE under Grants No. DE-FG02- 04ER46124.
%=================references===============================

%==================END============================

\end{document}